\documentclass{article}
\usepackage{adjustbox}
\usepackage{authblk}
\usepackage{pythonhighlight}
\usepackage{graphicx} % Required for inserting images
\usepackage{amsmath}
\usepackage{subcaption}
\usepackage{tabularx}
\usepackage{array}
\usepackage{float} 
\usepackage{biblatex}
\title{Exact and Efficient Numerical approaches to MIT Bag Model}

\author[1]{Amirhossein Rezaei\thanks{amirh.rezaei@mail.sbu.ac.ir}}
\author[1]{Mohammad Parsa Akrami\thanks{moh.akrami@mail.sbu.ac.ir}}

\affil[1]{\textit{Department of Physics, Shahid Beheshti University, Tehran, Iran}}

\date{March 2024}
\begin{document}
\maketitle
\begin{abstract}
    In this document, we examine exact and efficient numerical approaches to the MIT Bag Model, a theoretical framework used to describe the properties of bound quarks in Hadrons. We present the exact and Boundary Value Problem (BVP) numerical approaches. Both methods are effective in calculating the eigen-functions and energy levels. Notably, the precision of the BVP approach matches up to 10 decimal places when compared to the exact approach.
\end{abstract}

\section{Introduction}
The MIT Bag Model \cite{johnson1975} is a simple model used to describe the properties of bound quarks in Hadrons, without considering the strong interaction between the quarks. In this model the quarks which are described by the Dirac equation are confined within an enclosed area by introducing a scalar field $U_0$ outside the area of confinement and then taking the limit $U_0\to\infty$ which then gives us the MIT boundary condition that is written below :
\begin{equation}
    (1+i(\hat{n}.\Vec{\gamma}))\psi \bigg|_{Boundary}=0
\end{equation}
And $\Vec{\gamma}$ is:
\begin{equation}
    \Vec{\gamma}= \begin{pmatrix}
     0 & \Vec{\sigma} \\\\
    -\Vec{\sigma} & 0
    \end{pmatrix}
\end{equation}
We also have the Dirac equation ($(i\gamma^\mu\partial_\mu-m)\Psi$).Using this equation with the boundary condition above we can find the bound state energy levels of our confined quarks.
\section{Spherical Symmetry}
First we want to solve the problem for confinement in a sphere.Spherical symmetry simplifies the problem and we can separate the radial part from angular part in this way:
\begin{equation}
\Psi^k_{j j_3} = \begin{pmatrix}
    
g_\kappa(r)y_{j l_A}^{j_3}  \\\\
if_\kappa(r)y_{j l_B}^{j_3}
\end{pmatrix}
\end{equation}
If we write our solutions in this way, our boundary condition gets simplified to  $f_{\kappa}(R)+g_{\kappa}(R)=0$. And we have these two differential equations describing $f_{\kappa}(r),g_{\kappa}(r)$:
\begin{equation}
    \begin{cases}
        \frac{df_{\kappa}(r)}{dr}+\frac{1-\kappa}{r}f_{\kappa}(r)=(m-E)g_{\kappa}(r)\\\\
        \frac{dg_{\kappa}(r)}{dr}+\frac{1+\kappa}{r}g_{\kappa}(r)=(m+E)f_{\kappa}(r)
    \end{cases}
\end{equation}
\subsection{Analytic solution}
By solving this set of differential equations we find solutions in the forms of Spherical Bessel and Neumann functions which then we can proceed to discard the Neumann functions due to the fact that they diverge at $r\to0$.And then we reach these sets of answers for positive and negative $\kappa$ values:\\\\
for $\kappa<0:$
\begin{equation}
    \begin{cases}
        g_\kappa(r)=Nj_{|\kappa|-1}(Pr)\\\\
        f-\kappa(r)=-\frac{NP}{m+E}j_{|\kappa|}(Pr)
    \end{cases}
\end{equation}
for $\kappa>0:$
\begin{equation}
    \begin{cases}
        g_\kappa(r)=N'j_{\kappa}(Pr)\\\\
        f_\kappa(r)=\frac{N'P}{m+E}j_{\kappa-1}(Pr)
    \end{cases}
\end{equation}
Where $P=\sqrt{E^2-m^2}$\\\\
For $s_{\frac{1}{2}}$ states ($\kappa=-1:$) we have :
\begin{equation}
    \begin{cases}
        g_{-1}(r)=NJ_0(Pr)=N\frac{Sin(Pr)}{Pr}\\\\
        f_{-1}(r)=-\frac{NP}{m+E}j_1(Pr)=-\frac{NP}{m+E}(\frac{Sin(Pr)}{Pr}-\frac{Cos(Pr}{Pr})
    \end{cases}
\end{equation}
Applying normalization ($\int_{0}^{R} [f_{\kappa}^2+g_{\kappa}^2] r^2 \, dr$):
\begin{equation}
    N=\bigg[\frac{2PR-Sin(2PR)}{4P^3}+\frac{2P^2R^2+PRSin(2PR)+2Cos(2PR)-2}{4(m+E)^2P^4R}\bigg]^{-\frac{1}{2}}
\end{equation}
Applying the boundary condition :
\begin{equation}
    \frac{Cot(PR)}{(m+E)R}-\frac{1}{(m+E)PR^2}+\frac{1}{PR}=0
\end{equation}

\subsection{Boundary Value Problem (BVP)}

In this work, we tackled a boundary value problem (BVP) for the Dirac equation with MIT boundary conditions. We started by defining the boundary conditions in the function \texttt{bc(ya, yb)}, which takes the solutions at the start and end of the interval as input and returns the residuals of the boundary conditions. To solve the BVP, we created a mesh in the interval [0, 1] using the \\ \texttt{np.linspace(1e-6, 1, 1000)} function from NumPy. This function returns evenly spaced numbers over a specified interval, providing the necessary structure for the BVP solver \cite{2020SciPy-NMeth}. The objective function for the optimization problem was defined in the function \texttt{objective(E\_)}. This function solves the BVP for a given energy \(E\) and returns the sum of the squares of the solutions, calculated as 

\begin{equation} 
\sum_{i}^{}{(f(x_i)^2 + g(x_i)^2)} x_i^2
\end{equation}

\noindent where each \(i\) is a data point. 

To find the energy that minimizes the objective function, we used the \\\texttt{differential\_evolution} function from the \texttt{scipy.optimize} module \cite{2020SciPy-NMeth}. This function implements the differential evolution algorithm, a type of evolutionary algorithm that is useful for multidimensional global optimization problems. We performed this optimization by searching in different energy intervals, which helped us avoid falling repeatedly into the ground state. After performing the optimization, we obtained a set of energy values that minimized the objective function. These energy values correspond to the eigenvalues of the Dirac equation under the given boundary conditions. After obtaining the energy that minimizes the objective function, we used the optimal ($f$) and ($g$) corresponding to this minimum energy (E). These optimal solutions represent the wave functions of the quantum system. Finally, we visualized the wave functions using the \texttt{matplotlib.pyplot} module. The plots provided a clear representation of the quantum states of the system under the given boundary conditions. This approach allowed us to solve the Dirac equation for a quantum system with MIT boundary conditions, providing valuable insights into the behavior of the system. Future work could extend this approach to other types of boundary conditions or quantum systems.

\subsection{Results}
Here, Figures 1 through 4 correspond to the Exact Approach, while Figures 5 and 6 are associated with the BVP Approach. (Note that the wave functions are not normalized.)
Additionally, Table 1 provides a comparison of the energy values of the solutions, which are equal up to approximately 10 decimal places.
\begin{figure}[H]
    \centering
    \includegraphics[width=0.8\textwidth]{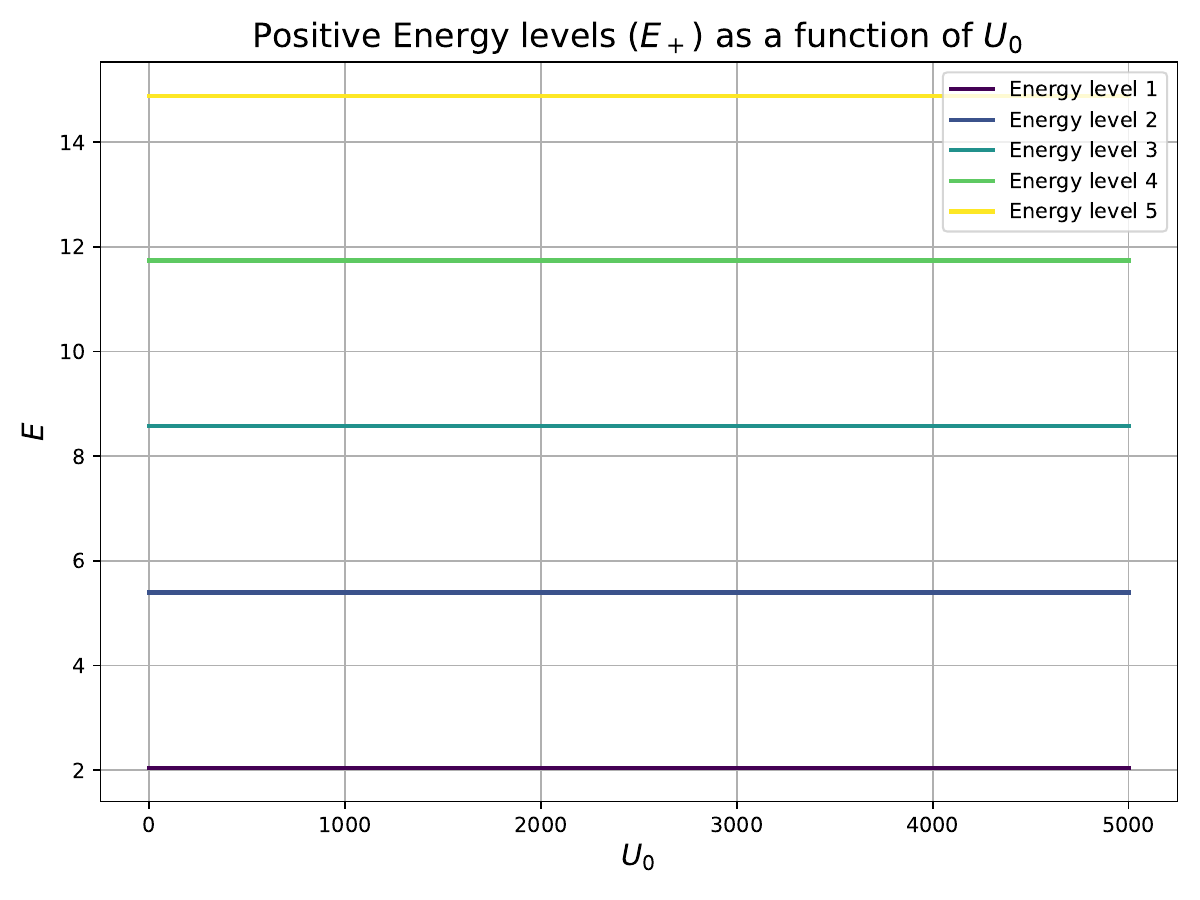}
    \caption{Positive Energy levels}
    \label{fig:positive_energy}
    
    \centering
    \includegraphics[width=0.8\textwidth]{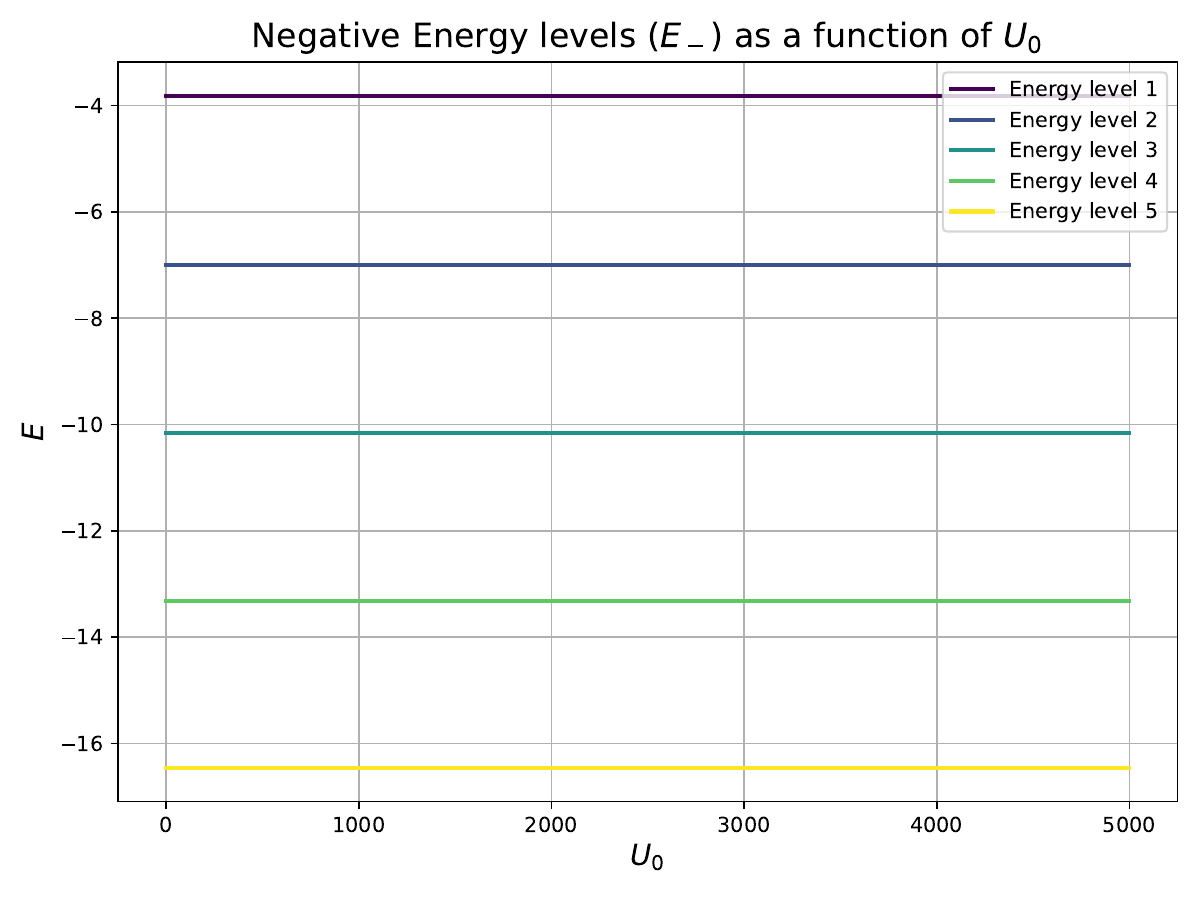}
    \caption{Negative Energy levels}
    \label{fig:negative_energy}
\end{figure}

\begin{figure}[H]
    \centering
    \includegraphics[width=0.8\textwidth]{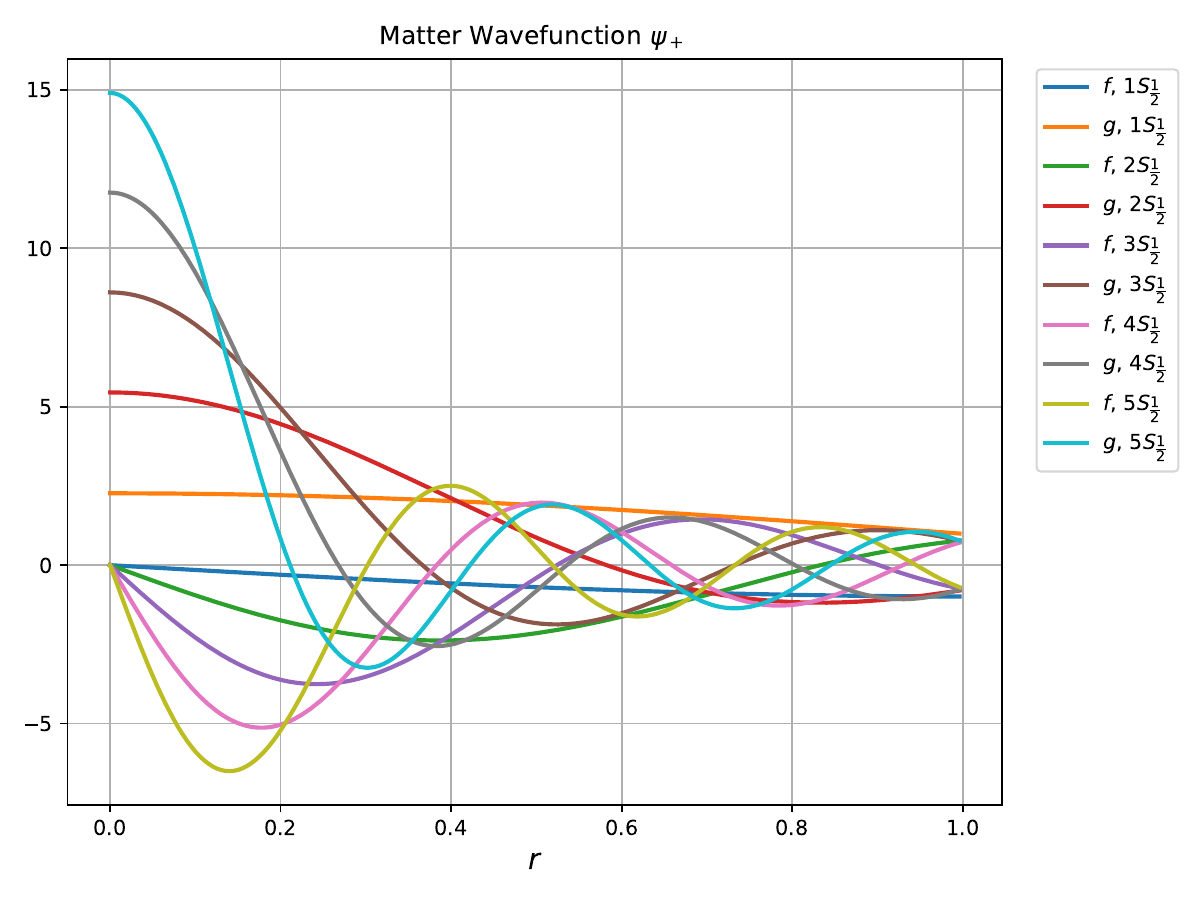}
    \caption{Matter Wave function, Exact}
    \centering
    \includegraphics[width=0.8\textwidth]{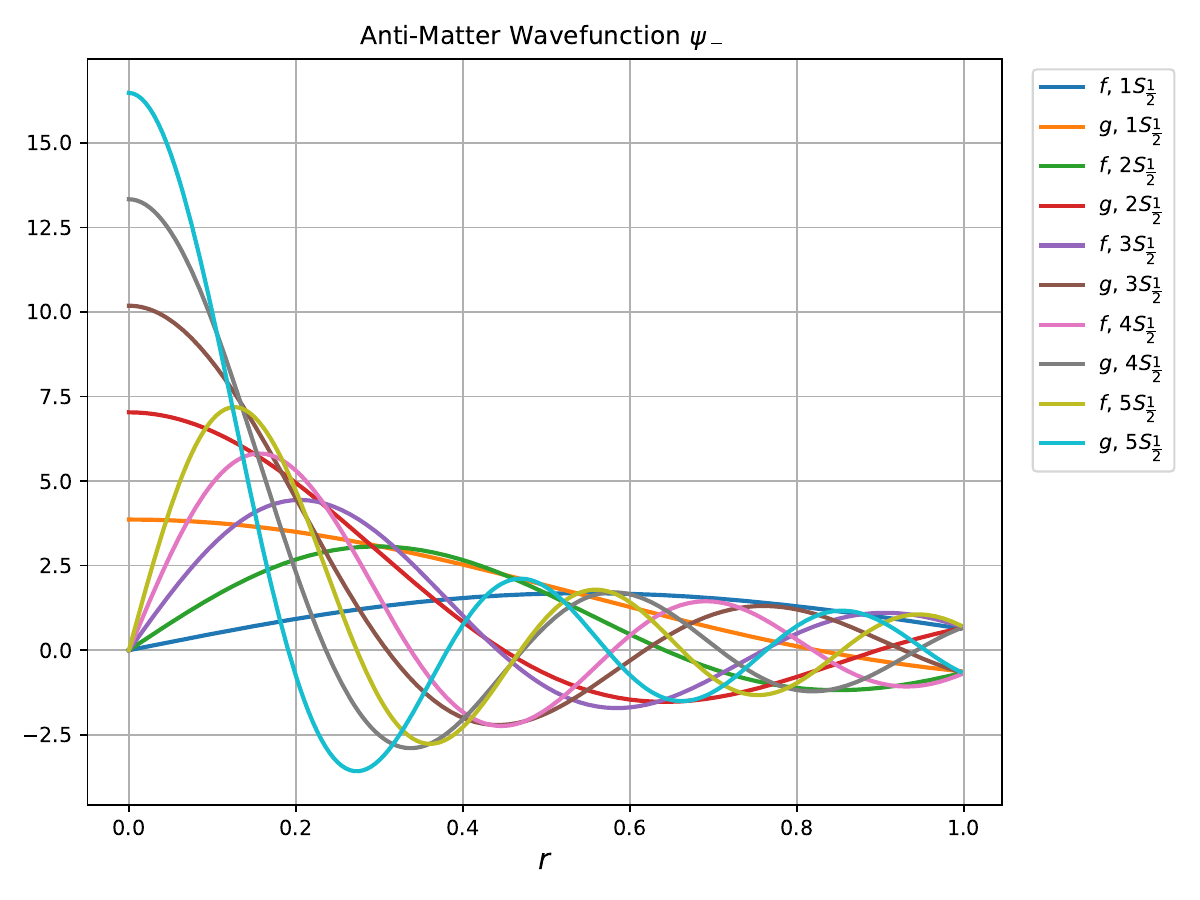}
    \caption{Anti Matter Wave function, Exact}
    \label{fig:matter and anti_matter}
\end{figure}

\begin{figure}[H]
    \centering
    \includegraphics[width=0.8\textwidth]{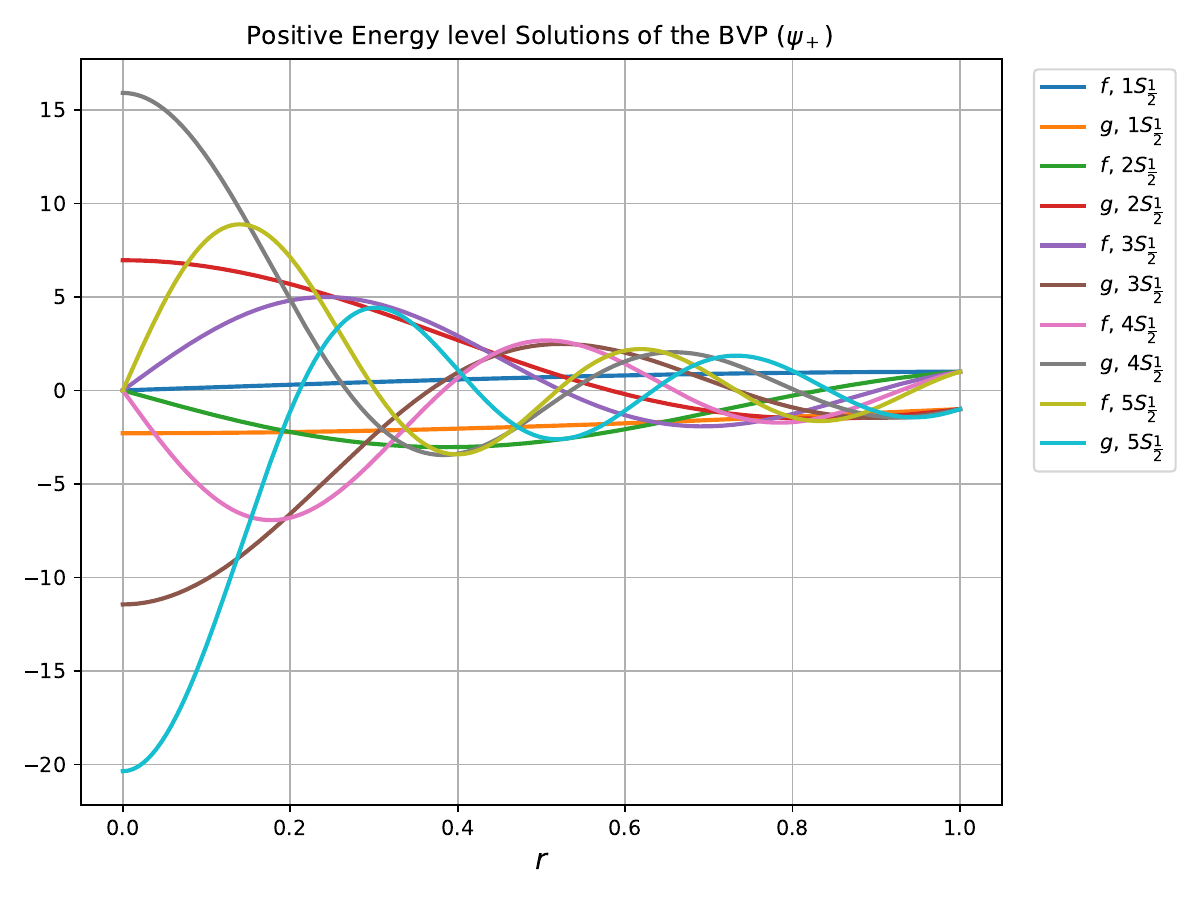}
    \caption{Matter Wave function, Numerical approach (BVP)}
    \centering
    \includegraphics[width=0.8\textwidth]{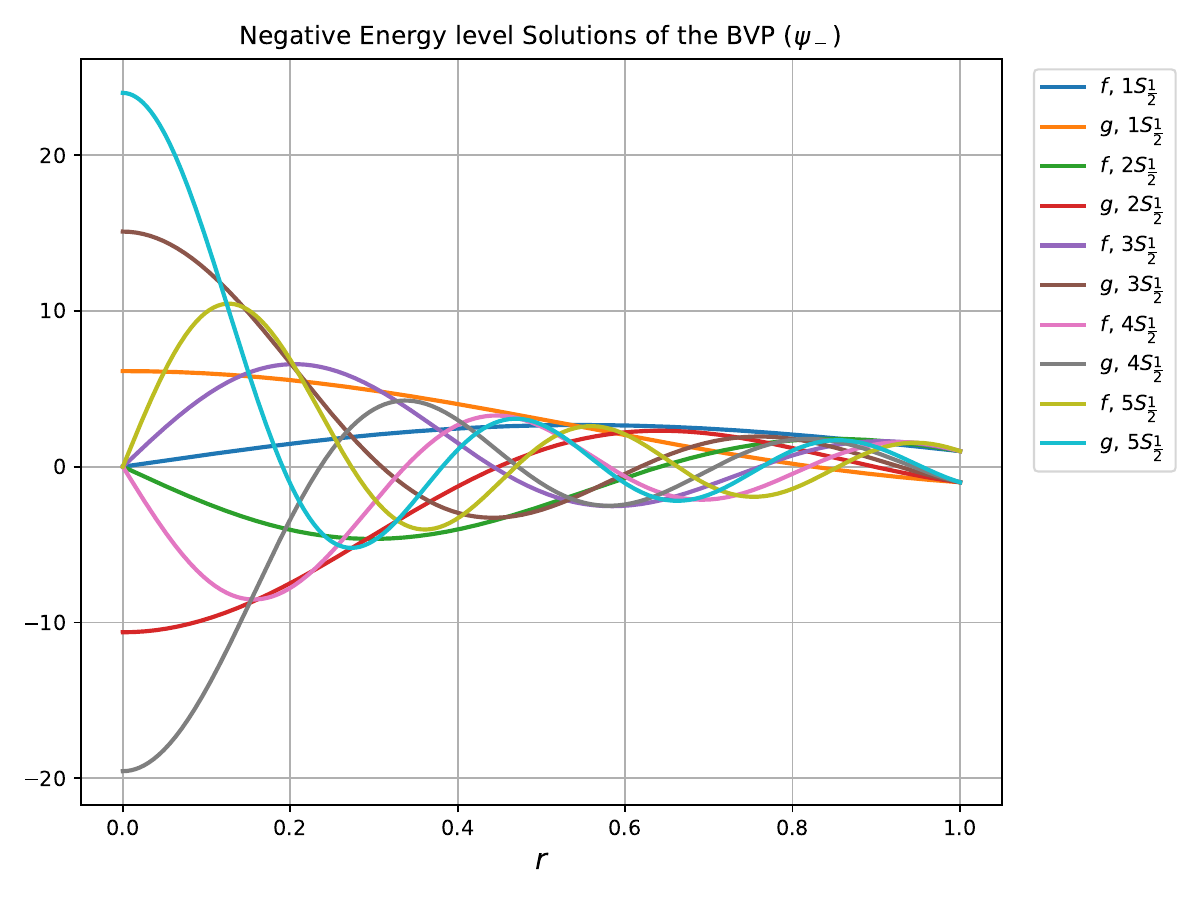}
    \caption{Anti Matter Wave function, Numerical approach (BVP)}
    \label{fig:anti_matter and negative_energy}
\end{figure}

\begin{table}[H]
\begin{adjustbox}{center}
\centering

\label{tab:my_label}
\resizebox{1.3\textwidth}{!}{%

{
\centering
\begin{tabular}{|c|c|c|c|}
\hline
\textbf{BVP Negative Energies} & \textbf{BVP Positive Energies} & \textbf{Exact Negative Energies} & \textbf{Exact Positive Energies} \\ 
\hline
-3.811538647779564 & 2.0427869427384784 & -3.811538647779367 & 2.042786942738485 \\
-7.002033295718347 & 5.396016117857496 & -7.002033295713394 & 5.396016117856196 \\
-10.163320735152427 & 8.57755878463046 & -10.163320735093786 & 8.577558784609673 \\
-13.315593577136053 & 11.736503959487846 & -13.315593576838708 & 11.736503959347777 \\
-16.463895601050194 & 14.887827486298722 & -16.463895600057722 & 14.887827485735363 \\
\hline

\end{tabular}%
}
 \par 
}

\centering
\end{adjustbox}
\caption{Calculated values of Positive and Negative Energies with BVP and exact approaches}
\end{table}

\section{Appendix}
\subsection{Exact Solution Code}
\begin{python}[language=Python]
import numpy as np
import matplotlib.pyplot as plt
from scipy.optimize import brenth
from tqdm import tqdm
import warnings

warnings.filterwarnings('ignore')

def const_q(E,m,U0):
    q = np.sqrt((m+U0)**2 - E**2)
    return q

def const_p(E,m):
    p = np.sqrt(E**2-m**2)
    return p

def const_N(R,E,m,U0):
    p, q = const_p(E, m), const_q(E, m, U0)
    return (R/(2*p**2) + R/(2*(m+E)**2) + (np.sin(2*p*R)/(4 * p**3)) * (p**2/((m+E)**2) - 1) - np.sin(p*R)**2 / (p**2 * R * (m+E)**2) + (np.sin(p*R)**2 / (2*p**2)) * (1/q + (q+2/R)/((m+E+U0)**2)))**-0.5

def const_M(R,E,m,U0):
    p, q = const_p(E, m), const_q(E,m, U0)
    return const_N(R,E,m,U0) * q * np.exp(q*R) * np.sin(p*R) / p

def solution(r, R, E, m, U0):
    p, q = const_p(E, m), const_q(E,m, U0)
    M, N = const_M(R,E,m,U0), const_N(R,E,m,U0)

    f = np.zeros_like(r)
    g = np.zeros_like(r)

    mask1 = r <= R
    mask2 = r > R

    r1 = r[mask1]
    r2 = r[mask2]

    g[mask1] = N * np.sin(p*r1)/ (p*r1)
    f[mask1] = -N * p * (np.sin(p*r1)/ ((p*r1)**2) - np.cos(p*r1)/ (p*r1)) / (m+E)

    g[mask2] = M*np.exp(-q*r2)/(q*r2)
    f[mask2] = -M*q*np.exp(-q*r2) * (1/(q*r2) + 1/(q*r2)**2) / (m+E+U0)

    return f, g

def solve_E(m, R, U0, E_range):
    def f(E, inf_limit = True):
        try:
            if not inf_limit:
                if m + U0 - E > 0:
                    term0 = np.tan(R * np.sqrt(E**2-m**2))
                    
                    if np.isclose(term0, 0):
                        return np.inf
                    
                    term1 = np.sqrt((E-m)/(E + m)) / term0
                    term2 = - 1/(R * (E + m))
                    term3 = np.sqrt((m + U0 - E)/(m + U0 + E))
                    term4 = 1/ (R * (m + U0 + E))
                    return term1 + term2 + term3 + term4

                else:
                    return np.nan
            
            elif inf_limit:
                return 1/(np.tan(const_p(E,m)*R) * ((m+E)*R)) - 1/((m+E)*const_p(E,m)*R**2) + 1/(const_p(E, m)*R)

        except ValueError:
            return np.inf

    roots = []

    for i in range(len(E_range) - 1):
        try:
            root = brenth(f, E_range[i], E_range[i + 1])
            roots.append(root)
        except ValueError:
            pass

    return np.array(roots)

def compute_E_values(m, R, U0_values, E_range, N = 10):
    E_values = [solve_E(m, R, U0, E_range) for U0 in tqdm(U0_values)]
    E_values_all = []
    start = 1 if np.all(E_range < 0) else 0

    for i in range(start, 2*N, 2):
        E_values_all.append([E[i] if len(E) > i else np.nan for E in E_values])
    return np.array(E_values_all)

def plot_energy_levels(U0_values, E_values_all, title, save_name):
    plt.figure(figsize=(8, 6))
    colormap = plt.cm.get_cmap('viridis', len(E_values_all))
    for i in range(len(E_values_all)):
        plt.plot(U0_values, E_values_all[i], label=f'Energy level {i+1}', linewidth=2, color=colormap(i))
    plt.legend()

    plt.xlabel(r'$U_0$', fontsize=14)
    plt.ylabel(r'$E$', fontsize=14)
    plt.grid(True)
    plt.title(title, fontsize=16)  
    plt.tight_layout()
    plt.savefig(save_name)  

    plt.show()

def plot_wavefunctions(r, U0_values, E_values_all, title, save_name):
    plt.figure(figsize=(8, 6))
    for energy_level in range(len(E_values_all)):
        f, g = solution(r, R, E_values_all[energy_level, -1], m, U0=U0_values[-1])
        plt.plot(r, f, linewidth=2, label=r'$f$'+f', {energy_level+1}'+ r'$S_\dfrac{1}{2}$')
        plt.plot(r, g, linewidth=2, label=r'$g$'+f', {energy_level+1}'+ r'$S_\dfrac{1}{2}$')
    plt.title(title)

    plt.legend(bbox_to_anchor=(1.2, 1.0), loc='upper right')
    plt.grid(True)
    plt.xlabel(r'$r$', fontsize=14)
    plt.tight_layout()
    plt.savefig(save_name)

    plt.show()

m = 0
R = 1

U0_values = np.linspace(0, 5000, 100) 
r = np.linspace(1e-5, 5, 1000)

E_range_negative = np.linspace(-1, -30, 100)
E_range_positive = np.linspace(1, 30, 100)

E_values_all = compute_E_values(m, R, U0_values, E_range_positive, N = 5)
np.savetxt('Results/exact_positive_energies.txt', E_values_all[:,-1], fmt= '%.15f')
plot_energy_levels(U0_values, E_values_all, title=r'Positive Energy levels ($E_+$) as a function of $U_0$', save_name = 'Results/Positive_Energy_levels.pdf')
plot_wavefunctions(r, U0_values, E_values_all, title = r'Matter Wavefunction $\psi_+$', save_name = 'Results/Matter_Wavefunction.pdf')

E_values_all = compute_E_values(m, R, U0_values, E_range_negative, N = 5)
np.savetxt('Results/exact_negative_energies.txt', E_values_all[:,-1], fmt= '%.15f')
plot_energy_levels(U0_values, E_values_all, title=r'Negative Energy levels ($E_-$) as a function of $U_0$', save_name = 'Results/Negative_Energy_levels.pdf')
plot_wavefunctions(r, U0_values, E_values_all, title = r'Anti-Matter Wavefunction $\psi_-$', save_name = 'Results/Anti_Matter_Wavefunction.pdf')
\end{python}

\subsection{BVP Solution Code}

\begin{python}[language=Python]
import numpy as np
from scipy.integrate import solve_bvp
import matplotlib.pyplot as plt
from scipy.optimize import differential_evolution

# Define the function for the differential equation system
def fun(x, y):
    k = -1
    m = 0
    dfdr = y[0]*(k-1)/x + (m - E)*y[1]
    dgdr = (-k-1)/x * y[1] + (m+E) * y[0]
    return np.vstack((dfdr, dgdr))

# Define the boundary condition residuals
def bc(ya, yb):
    return np.array([yb[0]-1, yb[1]+1])

# Create a mesh in the interval [0, 1]
x = np.linspace(1e-6, 1, 1000)

# Initial guess
y_init_1 = np.full((2, x.size), 3)

# Define the objective function for minimization
def objective(E_):
    global E
    E = E_[0]
    # Solve the boundary value problem
    res = solve_bvp(fun, bc, x, y_init_1, tol = 1e-5, max_nodes=1800)
    f = res.sol(x)[0]
    g = res.sol(x)[1]
    rho = np.dot(f**2+g**2, x**2)
    return rho

for pos in [False, True]:
    if pos:
        bounds_list = [(1, 4), (4, 8), (8, 10), (10, 12), (12, 15)]  # Add more bounds as needed
    else:
        bounds_list = [(-4, -1), (-8, -4), (-12, -8), (-16, -12), (-17, -15)]  # Add more bounds as needed

    optimal_E_vals = []

    plt.figure(figsize=(8, 6))

    for energy_level, bounds in enumerate(bounds_list):
        # Perform the differential evolution
        result = differential_evolution(objective, [bounds], tol = 1e-4)

        # Solve the boundary value problem with optimal E
        E = result.x[0]
        res = solve_bvp(fun, bc, x, y_init_1, tol = 1e-5)
        optimal_f = res.sol(x)[0]
        optimal_g = res.sol(x)[1]

        print(f"The optimal E for bounds {bounds} is {E} with a rho of {result.fun}")
        optimal_E_vals.append(E)

        # Plot the solution
        plt.plot(x, optimal_f, linewidth=2, label=r'$f$'+f', {energy_level+1}'+ r'$S_\dfrac{1}{2}$')
        plt.plot(x, optimal_g, linewidth=2, label=r'$g$'+f', {energy_level+1}'+ r'$S_\dfrac{1}{2}$')

    plt.legend(bbox_to_anchor=(1.2, 1.0), loc='upper right')

    plt.xlabel(r'$r$', fontsize=14)
    plt.grid(True)

    if pos:
        plt.title(r'Positive Energy level Solutions of the BVP ($\psi_+)$')
        plt.tight_layout()
        plt.savefig('Results/BVP_positive.pdf')
        np.savetxt('Results/BVP_positive_energies.txt', optimal_E_vals, fmt= '%.15f')
    else:
        plt.title(r'Negative Energy level Solutions of the BVP ($\psi_-)$')
        plt.tight_layout()
        plt.savefig('Results/BVP_negative.pdf')
        np.savetxt('Results/BVP_negative_energies.txt', optimal_E_vals, fmt= '%.15f')

    plt.show()

\end{python}

\printbibliography

@ARTICLE{johnson1975,
    author = {Johnson, K.},
    title = {The MIT bag model},
    journal = {Acta Phys. Pol. B},
    volume = {6},
    number = {12},
    pages = {8},
    year = {1975}
}

@ARTICLE{2020SciPy-NMeth,
  author  = {Virtanen, Pauli and Gommers, Ralf and Oliphant, Travis E. and
            Haberland, Matt and Reddy, Tyler and Cournapeau, David and
            Burovski, Evgeni and Peterson, Pearu and Weckesser, Warren and
            Bright, Jonathan and {van der Walt}, St{\'e}fan J. and
            Brett, Matthew and Wilson, Joshua and Millman, K. Jarrod and
            Mayorov, Nikolay and Nelson, Andrew R. J. and Jones, Eric and
            Kern, Robert and Larson, Eric and Carey, C J and
            Polat, {\.I}lhan and Feng, Yu and Moore, Eric W. and
            {VanderPlas}, Jake and Laxalde, Denis and Perktold, Josef and
            Cimrman, Robert and Henriksen, Ian and Quintero, E. A. and
            Harris, Charles R. and Archibald, Anne M. and
            Ribeiro, Ant{\^o}nio H. and Pedregosa, Fabian and
            {van Mulbregt}, Paul and {SciPy 1.0 Contributors}},
  title   = {{{SciPy} 1.0: Fundamental Algorithms for Scientific
            Computing in Python}},
  journal = {Nature Methods},
  year    = {2020},
  volume  = {17},
  pages   = {261--272},
  adsurl  = {https://rdcu.be/b08Wh},
  doi     = {10.1038/s41592-019-0686-2},
}

\end{document}